\documentclass[aps,prl,showpacs,twocolumn,superscriptaddress, floatfix, tightenlines, amsmath, amssymb]{revtex4-1}
\usepackage[pdftex]{graphicx} 

\graphicspath{{figures/}}
\usepackage{natbib}
\usepackage{booktabs}
\usepackage{xcolor}
\usepackage{url}
\usepackage{amsmath}
\usepackage[colorlinks=true, citecolor=blue, linkcolor=blue, urlcolor=blue]{hyperref}

\newcommand{\SrCa}{Sr$_3$CaIr$_2$O$_9$} 
\newcommand{\Na}{NaIrO$_3$}

\newcommand{\eloss}{$E_{\mathrm{loss}}$}

\begin{document}

\title{On the electronic ground state of two non-magnetic pentavalent honeycomb iridates}
\author{A. de la Torre}
\affiliation{Department of Physics, Brown University, Providence, Rhode Island 02912, United States}
\author{B. Zager}
\affiliation{Department of Physics, Brown University, Providence, Rhode Island 02912, United States}
\author{J. R. Chamorro}
\affiliation{Department of Chemistry, The Johns Hopkins University, Baltimore, MD, USA}
\affiliation{ Institute for Quantum Matter, William H. Miller III Department of Physics and Astronomy, Johns Hopkins University, Baltimore, MD, USA}
\author{M. H. Upton}
\affiliation{Advanced Photon Source, Argonne National Laboratory, Argonne, Illinois 60439, USA}
\author{G. Fabbris}
\affiliation{Advanced Photon Source, Argonne National Laboratory, Argonne, Illinois 60439, USA}
\author{D. Haskel}
\affiliation{Advanced Photon Source, Argonne National Laboratory, Argonne, Illinois 60439, USA}
\author{D. Casa}
\affiliation{Advanced Photon Source, Argonne National Laboratory, Argonne, Illinois 60439, USA}
\author{T. M. McQueen}
\affiliation{Department of Chemistry, The Johns Hopkins University, Baltimore, MD, USA}
\affiliation{ Institute for Quantum Matter, William H. Miller III Department of Physics and Astronomy, Johns Hopkins University, Baltimore, MD, USA}
\affiliation{Department of Materials Science and Engineering, The Johns Hopkins University, Baltimore, MD, USA}
\author{K. W. Plumb}
\affiliation{Department of Physics, Brown University, Providence, Rhode Island 02912, United States}
\date{\today}

\begin{abstract}

We investigate the electronic structure of two Ir$^{5+}$ honeycomb iridates, \SrCa{} and \Na{}, by means of resonant x-ray techniques. We confirm that \SrCa{} realizes a large spin-orbit driven non-magnetic $J = 0$ singlet ground state despite sizable tetragonal distortions of Ir coordinating octahedra. On the other hand, the resonant inelastic x-ray spectra of \Na{} are drastically different from expectations for a  Mott insulator with octahedrally coordinated Ir$^{5+}$. We find that the data for \Na{} can be best interpreted as originating from a narrow gap non-magnetic $S = 0$ band insulating ground state. Our results highlight the complex role of the ligand environment in the electronic structure of honeycomb iridates and the essential role of x-ray spectroscopy to characterize electronic ground states of insulating materials.

\end{abstract}
\maketitle
\section{Introduction}

\begin{figure}[!t]
  \begin{center}
    \includegraphics{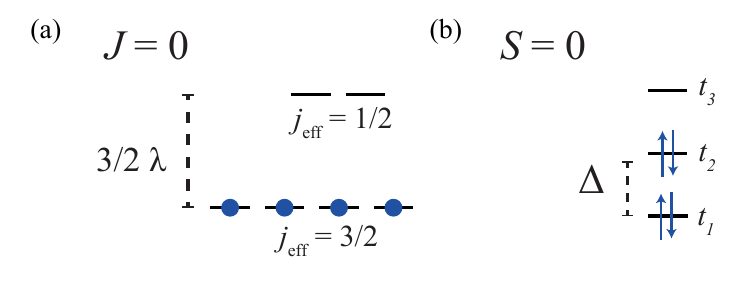}
  \caption{Schematic of two limiting cases for an octahedrally coordinated  Ir$^{5+}$ ion: (a)  $J = 0$ state and  (b) $S = 0$, with $\lambda >> \Delta$ and $\Delta \approx \lambda \approx J_H$, respectively.}
  \label{fig:Model}
  \end{center}
\end{figure}

New emergent correlated phenomena have been predicted to arise in materials with $4/5d$ transition metals in where spin-obit coupling ($\lambda$) and electron-electron correlations ($U$) are the dominant energy scales, including topologically protected magnetic states and complex magnetic orders. The essential physics occurs at the single ion level. Operating within the $j-j$ coupling limit, the interplay of cubic crystal field and spin orbit coupling acts to generate $j_{\rm eff} = 3/2$ quartet and a $j_{\rm eff} = 1/2$ doublet states from the $d$ electron levels \cite{Novel_Jeff_iridates}. Depending on the electron filling, the $j_{\rm eff}$ states form the basis for many novel quantum phases including Kitaev spin liquids \cite{Jackeli:2009}, spin-orbital liquids \cite{Chen:2010, Yamada:2018, Natori:2018}, multipolar phases \cite{Chen:2010}, and other unconventional ordered states \cite{Erickson:2007, Witczak-Krempa, Rau_AnnRevCondMatt}. The case of $d^4$ electron filling and octahedral coordination is of particular interest, as it realizes a nominally non-magnetic $J_{\rm eff} = 0, m_{J_{\rm eff}} = 0$ single-ion singlet \cite{Cao_2018_review,Bhowal_2021,PhysRevB.93.035129,arxiv_2206_05223}, as shown in Fig.~\ref{fig:Model} (a). This non-magnetic Mott insulator state has attracted recent attention as a platform to study excitonic magnetism. Although there are no preformed local moments on $J \!=\!0$  ions, if the superexchange energy is comparable to the atomic  spin-orbit coupling strength the $J\!=\!0$ ground state and first excited $J\!=\!1$ triplet can mix. Exchange driven condensation of the $J\!=1\!$ excited states results in an excitonic magnetic order from which novel magnetic phases can arise \cite{Khaliullin_excitonic,Svoboda_model_magnetic_interactions,PhysRevB.91.054412}. Such condensate can lead to quantum spin liquid physics when realized on a honeycomb lattice \cite{Kitaev_QSL_3/2}. 

Two recently synthesized pentavalent honeycomb iridates \SrCa{} and \Na{} display similar nearly  temperature independent magnetic susceptibilities with small paramagnetic moments \cite{C5DT03188E} that may indicate non-magnetic $J_{\rm eff}\!=\!0$ Mott insulating ground states. \SrCa{}, is a 2:1 ordered perovskite isostructural to Sr$_3$CaRu$_2$O$_9$ (space group $P21/c$) with a buckled honeycomb lattice and corner-sharing IrO$_6$ octahedra \cite{SrCa_Wilson}. Structural refinements for powder x-ray and neutron diffraction indicate that there are two non-equivalent IrO$_6$ octahedra in the structure of \SrCa{}, each having a different degree of cubic symmetry breaking, with a broad distribution of Ir-O bond distances. While the overall structure of the first Ir site, Ir$_A$, is characterized by  "three-long, three-short" Ir-O bond pattern (average Ir-O$_{\rm min}$/Ir-O$_{\rm max}$ = 0.86), with larger tetragonal distortions than that of Sr$_3$CaRu$_2$O$_9$ \cite{Rijssenbeek2002}, the second Ir site, Ir$_B$, is only weakly distorted (average Ir-O$_{\rm min}$/Ir-O$_{\rm max}$ = 0.92). Curie-Weiss analysis of the magnetic susceptibility finds a Weiss temperature  of $\theta_{\rm CW}\! =\! -8(5)$~K, and a small paramagnetic moment ($\mu^{\rm Sr}_{\rm eff} = 0.18~\mu_B/{\rm Ir}$) consistent with a $J\! =\! 0$ ground state.

On the other hand, \Na{} is synthesized via  oxidative Na-deintercalation of Na$_2$IrO$_3$ \cite{PhysRevB.82.064412} and maintains the edge-sharing IrO$_6$ honeycomb lattice of the parent compound with an average $d_{\rm{Ir-Ir}} \! \approx \! 3 {\rm \AA}$. Na-Ir stacking disorder in \Na{} reduces the space group to $P\bar{1}$. Similarly to \SrCa{}, Curie-Weiss analysis of the weakly temperature dependent magnetic susceptibility in the temperature range $3 < T < 300$~K found small paramagnetic moment ($\mu^{\rm Na }_{\rm eff} = 0.18~\mu_B/{\rm Ir}$) and Curie-Weiss temperature of $\theta_{\rm CW} = 0$~K. This observation led to a description of \Na{} and \SrCa{} as the first realizations of  non-magnetic $J\! = \! 0$ Mott insulators on the honeycomb lattice.  Thus, these materials are potential candidates to explore excitonic magnetism, but an investigation of the electronic structure is first necessary in order to verify the $J\!=\!0$ nature in each compound.

The non-magnetic $J_{eff} = 0$ ground state in octahedrally coordinated Ir$^{5+}$ compounds is sensitive to structural details. For instance, local non-cubic distortions of IrO$_6$ octahedra ($\Delta$) can result in a non-zero magnetic moment. At the single-ion level, the $t_{2g}$ orbitals for octahedrally coordinated $Ir^{5+}$ can be described in two limiting cases of $j - j$ coupling, $\Delta\! <<\! \lambda, J\! =\! 0$, Fig.~\ref{fig:Model} (a), and Russel-Saunders coupling,  $\Delta\! >>\! \lambda$, $S \!=\! 1$. When the distortion-driven splitting of the $t_{2g}$ orbitals is comparable to the Hund's coupling in iridates, a non-magnetic $S \! =\! 0$ can be stabilized \cite{Breakdown_J0_band_structure} [Fig.~\ref{fig:Model} (b)]. Such a band insulating ground state with strong-spin orbit coupling is of interest in its own right because of the potential to realize flat bands and topological excitations \cite{PhysRevB.104.235202}. At intermediate couplings, $\Delta/\lambda\!\approx\! 1$, non-cubic crystal fields might lead to a $J \neq 0$ ground state, as characterized by the expectation value $\langle J^2 \rangle$ \cite{Bhowal_Hamiltonian_d4}. Additional departures from the single-ion limit can result from direct Ir-Ir hopping, $t$ \cite{BaZn_other6H_RIXS_Nag}. Interestingly, most of the investigated octahedrally coordinated pentavalent iridates fall close to the $j-j$ coupling limit, but have been characterized by a small non-zero total angular momentum \cite{Bhowal_Hamiltonian_d4, Breakdown_J0_band_structure, Bhowal_theory_double_perovskite_Ca_MMg,aczel2022spinorbit}, including double perovskites Sr$_2$YIrO$_6$ \cite{SrYIrO_RIXS, SrYIrO_thermo}, Ba$_2$YIrO$_6$ \cite{BaY_RIXS_Thermo_Nag}, and the family of $6H$-perovskites Ba$_3$XIr$_2$O$_9$ (X = Sr,Ca,Zn,Mg) \cite{BaZn_other6H_RIXS_Nag, BaZn_thermo_Nag}. Given that the many closely competing energy scales can lead to drastically different insulating ground states with the same zero net magnetic moment, it is essential to investigate these materials using techniques that are directly sensitive to Ir $5d$-orbital occupation and ligand environment. Such information can be obtained directly through X-ray absorption spectroscopy (XAS) and Resonant Inelastic X-ray Scattering (RIXS) measurements at the Ir $L_3$ absorption edge \cite{PhysRevB.84.020403,PhysRevB.86.195131,PhysRevLett.110.076402}. 

In this work, we use XAS and RIXS to study the interplay of energy scales and local structural distortions in the electronic structure of the two Ir$^{5+}$ honeycomb iridates \SrCa{} and \Na{}. The measured RIXS spectrum in \SrCa{} is consistent with a nearly ideal $J_{\rm eff}\!=\! 0$ despite the presence of tetragonally distortorted IrO$_6$ octahedra. We find that local disorder results in a distribution of crystal field splittings that broadens the RIXS spectra for \SrCa{} and allows us to extract information about the distribution of energy scales from this disorder. On the other hand, the RIXS spectrum of \Na{} is  qualitatively different from that of \SrCa{} and all other reports for Ir$^{5+}$ compounds. In particular, we do not observe a well defined excitation on the scale of $3/2 \lambda$. The spectra is instead comprised of a  broad continuum of excitations for energy  ($\Delta E$) above 500~meV. In attempt to account for this,  we first consider a single ion model for \Na{} and explore the effects of different non-cubic crystal fields and exchange splitting in the RIXS spectrum. Qualitative consistency between this Mott insulating scenario and the RIXS data requires large octahedral distortions and  non-zero magnetic moments that interact via an Ising exchange and antiferromagnetically order above $T\!=\!300$~K. Although not entirely ruled out by the existing data, magnetic order is unlikely in \Na{}. We then consider a second model in the limit of a flat-band narrow-gap $S\!=\!0$ insulator, in agreement with density functional theory calculations for a fully relaxed crystal structure. We find that such a band insulating state is the most likely description of the electronic state of \Na{}. Our result highlights the capabilities of resonant spectroscopic x-ray techniques to provide detailed information about the local environment and electronic structure of quantum materials and further emphasizes how the many competitive energy scales in honeycomb iridates act to produce intrinsically different magnetic and electronic ground states.

\section{Experimental details}
High quality powder samples of \SrCa{} and \Na{} were grown as described in Ref. \cite{C5DT03188E}. RIXS measurements were performed at Sector 27 (MERIX) at the Advanced Photon Source (APS) of the Argonne National Laboratory  \cite{SHVYDKO2013140}. A  2~m radius spherically diced Si(844) analyzer was used to provide an overall energy resolution (full width at half maximum, FWHM) of 33~meV. All measurements were performed with a fixed scattering angle of  $2\theta\! =\! 90^{\circ}$ in a specular geometry. Room temperature X-ray absorption spectroscopy (XAS) was performed on \SrCa{}, \Na{}, IrO, $\alpha$-Li$_2$IrO$_3$, and Ag$_3$LiIr$_2$O$_6$, in transmission at APS 4-ID-D and analyzed using the Larch \cite{Newville_2013} software packages. Exact diagonalization and simulations of RIXS cross-sections were carried out using the EDRIXS software package \cite{EDRIXS}. The DFT calculations for \Na{} were performed within the Perdew-Burke-Ernzernhof formulation of generalized gradient approximation (GGA) using the Quantum Espresso software package \cite{Giannozzi_2009_QE,Giannozzi_2017_QE}. Fully relativistic potentials were used for the spin-orbit calculations. For the fully relaxed calculations, the reported lattice parameters and corresponding atomic positions, assuming perfect Na stacking \cite{C5DT03188E}, are allowed to vary to achieve forces in each atom below $10^{-3}$~Ry/a.u. and pressure below 1~kbar. A $15\! \times \!15\!\times \! 15$ k-grid was used for density of states calculations. 

\section{Experimental Results}
\subsection{XAS}
\begin{figure}[!t]
  \begin{center}
    \includegraphics{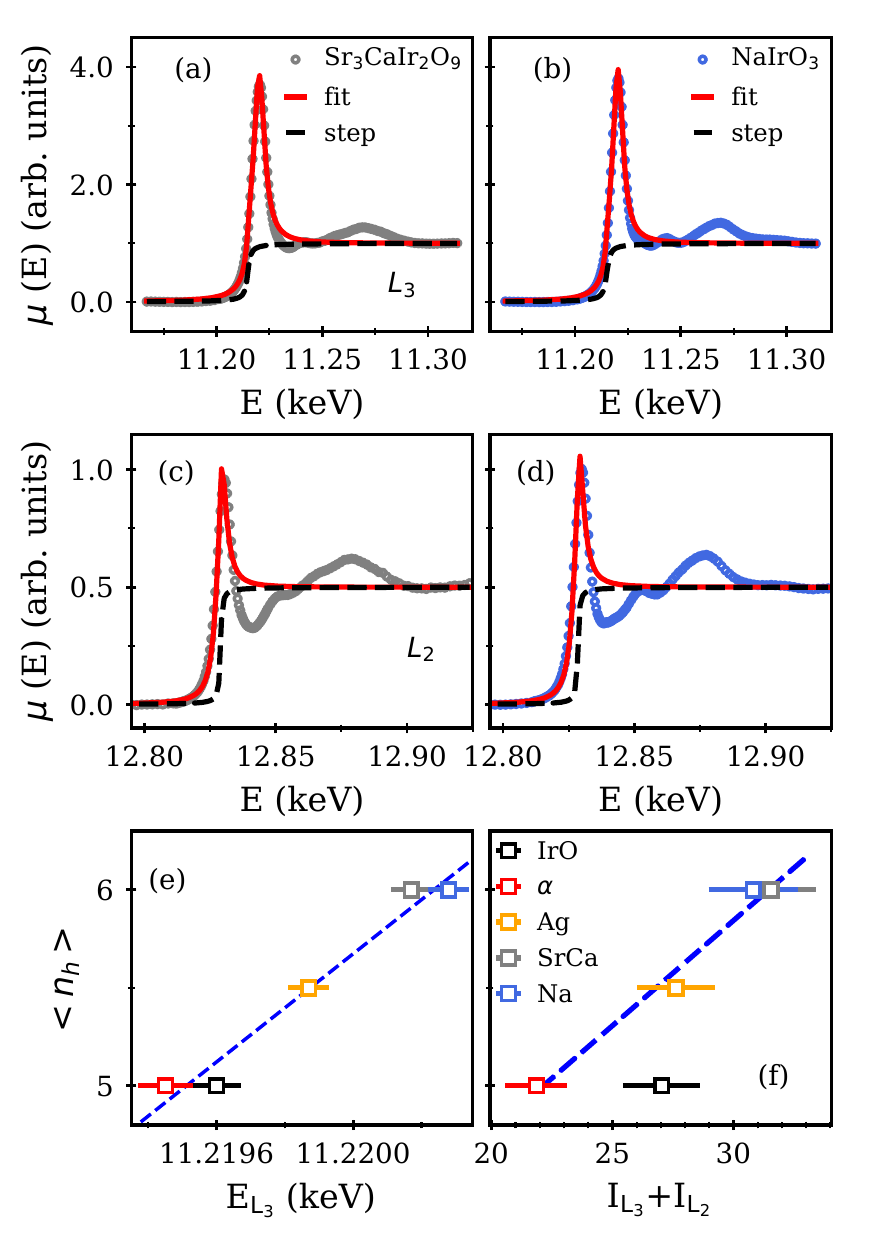}
  \caption{(a) Ir L$_3$ edge XAS intensity in \SrCa{} and ,(b), in \Na{}. (c) Ir L$_2$ edge XAS in \SrCa{} and, (d), in \Na{}. All data was taken at $T \!=\! 300$~K. Red line is a fit to a Lorentzian peak and an arctangent step (dotted black line), as explained in the main text. Direct comparison of the (e) $L_3$ white line and (f) $L_3 + L_2$ integrated intensity of \SrCa{} (SrCa) and \Na{} (Na) to that of IrO, $\alpha$-Li$_2$IrO$_3$ ($\alpha$) and Ag$_3$LiIr$_2$O$_6$ (Ag). Dotted blue line is a linear fit.}
  \label{fig:XAS}
  \end{center}
\end{figure}

We first examine the x-ray absorption spectra of \Na{} and \SrCa{} to provide an overall picture of their electronic structure and to verify the Ir$^{+5}$ oxidation state. This is a particularly important measurement for \Na{} because of the possibility for mixed Ir valence. While \Na{} powder phases with intermediate stoichiometry cannot be synthesized via \textit{chimie douce} oxidative deintercalation, it is possible that the oxidation state of Ir is altered upon irradiation with an intense x-ray beam \cite{C5DT03188E}. Moreover, single crystals of Na$_{0.7}$IrO$_3$ with mixed valence Ir ($\mu_{\rm{eff}} = 1.1~\mu_B/\rm{Ir}$) have been recently synthesized via flux methods \cite{NaxIrO3}. In Fig.~\ref{fig:XAS} (a),(c) we show the normalized absorption intensity for \SrCa{} at the Ir L$_3$ and L$_2$ edges, respectively. The equivalent data set for \Na{} is shown in Fig.~\ref{fig:XAS} (b),(d). 

In order to assess the Ir oxidation states, we fit the XAS white-line with a Lorentzian peak and arctangent step \cite{XAS_HighPressure_Ir327} and examine both the Ir ${\rm L}_3$ edge peak position, $E_{{\rm L}_3}$, [Fig.~\ref{fig:XAS} (e)] and the weighted sum of numerically integrated L$_3$ and L$_2$ white-line intensities $\langle n_h \rangle \propto {\rm L}_3 + (E_{{\rm L}_3}/E_{{\rm L}_2}) {\rm L}_2$ with \cite{PhysRevB.36.2972}, [Fig.~\ref{fig:XAS} (f)]. We also show the corresponding quantities determined from measurements on Ir$^{4+}$ standards IrO, $\alpha$-Li$_2$IrO$_3$ and an intermediate oxidation sate Ag$_3$LiIr$_2$O$_6$ \cite{PhysRevB.104.L100416,SM}. Such an analysis has successfully been used to determine the oxidation state of $5d$ compounds \cite{Clancy_5d_XAS,PhysRevB.91.214433, PhysRevB.104.L100416}. We find that the white line positions and integrated intensities are consistent with  Ir$^{5+}$ in both \SrCa{} and \Na{}. We also extract a branching ratio for \SrCa{} of $BR\! =\! I_{L_3}/I_{L_2}\!=\!5.8$ and for \Na{} $BR \! =\! 5.2$. These values are consistent with reports for other pentavalent iridates \cite{Clancy_5d_XAS}.

\subsection{RIXS on \SrCa{}}
\begin{figure}[!t]
  \begin{center}
    \includegraphics{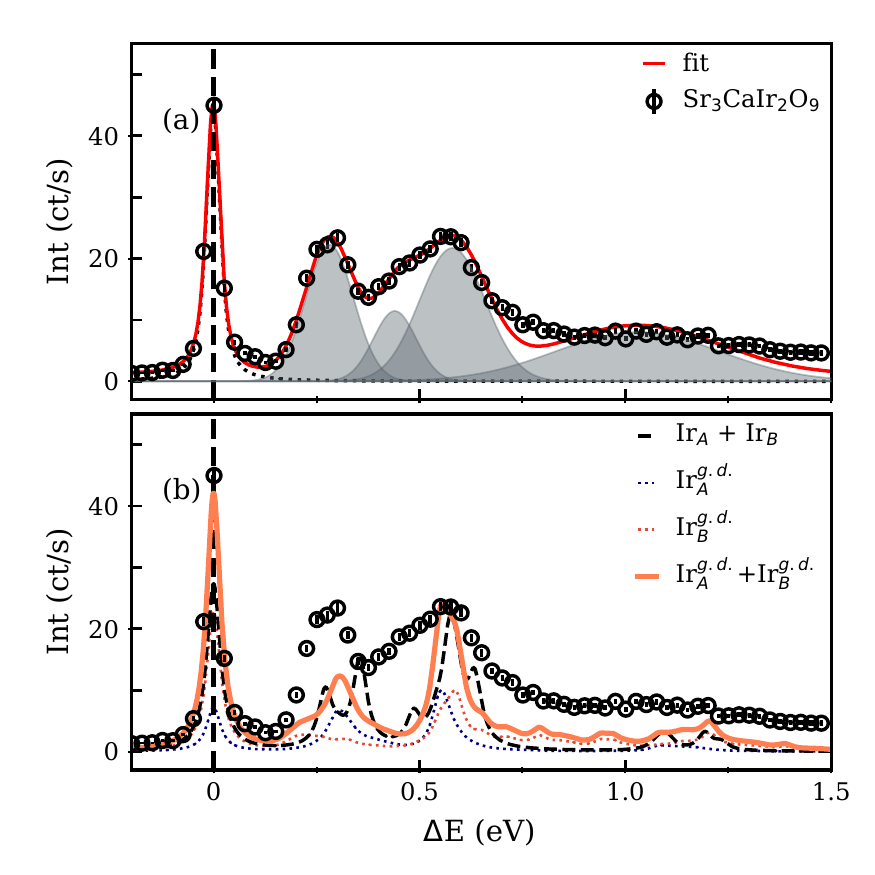}
  \caption{(a)  Ir L$_3$ RIXS intensity at $E_i \! = \! 11.217$ keV, in \SrCa{} at $T \!= \!20$~K. Vertical dotted line indicates \eloss{} $=\! 0$. Red line is a fit to a Voigt elastic line (dotted line) and four Gaussians peaks (grey shaded). (b) Calculated powder average RIXS intensity for two Ir$^{5+}$ sites with nonequivalent tetragonally distorted octahedral environments as described in the main text (black dashed line). The blue and red dotted lines show the corresponding RIXS intensities when including random tetragonal fields to account for static disorder. The total spectrum is shown in orange.}
  \label{fig:SrCA}
  \end{center}
\end{figure}

We use RIXS to reveal more detail of the electronic structure of \SrCa{}. In the limit of a large cubic crystal field, weak trigonal or tetragonal distortions, and small inter-site hopping, the RIXS spectrum for octahedrally coordinated Ir$^{5+}$is characterized by two low energy ($\Delta E\!< \! 1$~eV),  inelastic features  associated with intra-$t_{2g}$ excitations into a $J\! =\! 1$ triplet and a $J\! =\! 2$ quintuplet respectively \cite{Bhowal_Hamiltonian_d4} [Fig. \ref{fig:TS_diagram} (b)]. Most octahedrally coordinated Ir$^{5+}$ oxides show deviations from the large cubic crystal field limit. For example, in $6H$ perovskites \cite{SrYIrO_RIXS}, the small Ir-Ir ($d_{\rm Ir-Ir} \! \approx \! 2.75 ~ {\rm \AA}$) distances promote electron delocalization and give rise to additional inelastic features in the RIXS spectrum. In Fig.~\ref{fig:SrCA} (a), we show $T\!=\! 20$~K RIXS signal from powder samples of \SrCa{} at $E_i\! =\! 11.215$~keV for  $\Delta E \! < \!1.5$~eV.  It was necessary to include four Gaussian functions with $120<$ FWHM $<470$~meV and a resolution limited Voigt elastic line profile (FWHM = 33~meV) to fit the RIXS spectrum. Given the large average $d_{\rm{Ir-Ir}}\! =\! 3.98 (3)~{\rm \AA}$ in \SrCa{}, Ir-Ir hopping is not a likely possibility. Instead,  the four RIXS features can be consistently explained to arise from inter  $t_{2g}$ transitions on two nonequivalent Ir sites   and the presence of local structural disorder due to the broad distribution of Ir-O bond distances \cite{C5DT03188E}.

We consider a  model  with $U \! =\!  2$~eV, $J_H \! = \! 0.285$~eV, $\lambda \! = \! 0.36$~ eV and two non-interacting Ir sites labeled Ir$_A$ and Ir$_B$. The effect of a tetragonal distortion of IrO$_6$ octahedra is parameterized through  $\delta_1, \delta_2$ splitting of the  Ir $t_{2g}$ orbitals \cite{EDRIXS}. We allow for different degrees of tetragonal distortion on each site and find that the parameters $\delta^A_1 \!=\! -0.18 $~eV, $\delta^A_2 \!=\! - 0.03$~eV and $\delta^B_1 \!=\! -0.1 $~eV, $\delta^B_2 \!=\! - 0.05$~eV give reasonable agreement with the data. We include a final state lifetime of $\Gamma \!=\!0.02$~eV \cite{PhysRevB.104.L100416} and powder average the calculated RIXS cross-section to compare with our data. The total intensity is shown in Fig. \ref{fig:SrCA} (b) (black dashed line). While this model is in qualitative agreement with our data, it does not capture the considerable broadening of the $dd$ excitations. Attributing this broadening solely to electron delocalization would imply a minimum bandwidth of $\approx 120$~meV, which is not likely given that Ir - Ir hopping is expected to be negligible in in \SrCa{}. On the other hand, local structural disorder is a possibility that cannot be ruled out with the existing powder x-ray and neutron diffraction data, since these measurements are only sensitive to the average crystalline structure. Moreover, a broad distribution of Ir-O bond distances is known to occur in \SrCa{} \cite{C5DT03188E}. Such local structural disorder on the ligand could lead to a distribution of IrO$_6$ environments throughout the sample, resulting in an apparent broadening of crystal field excitations \cite{Mauws_2021_DisorderCF}. In order to incorporate such local disorder in RIXS spectra we consider a model that includes random tetragonal fields sampled from a normal distribution with mean values of $\delta_1, \delta_2$ and a standard deviation of  $\sigma \! = \! 0.2$~eV which provide the best agreement with the data (orange line). The RIXS calculation including crystal field disorder better recovers the overall energy broadening of the data as well as the intensity ratio between inelastic features, as shown in Fig. \ref{fig:SrCA}. The powder nature of our samples prevents us from ruling out or confirming additional broadening due to dispersive spin-orbit excitons, as seen in other $J \! = \! 0$ iridates \cite{arxiv_2206_04721}. This analysis highlights the utility of high resolution RIXS as a measure of local structural disorder in correlated insulators. Moreover, it allows us to directly extract an energy scale and a distribution for the local Ir-O disorder. Thus, RIXS measurements are highly complementary to local structural measurements such as Pair Distribution Function or Extended X-ray Absorption Fine Structure \cite{PhysRevB.104.L100416}.

Based on the mean crystal field energies from our RIXS model, we  compute an average expectation value for the total angular momentum, $\langle J^2 \rangle \! = \! 0.03$ \cite{EDRIXS}. Thus, our results confirm that \SrCa{} is very well described in the $j-j$ coupling limit of a $J_{eff}=0$ Mott insulator and that the weak moment extracted from susceptibility measurements is due to magnetic impurities \cite{C5DT03188E,SrCa_Wilson}. However, the large ~250~meV energy gap makes magnetic exciton condensation via application of hydrostatic pressure, carrier doping or other means an unlikely possibility in this material \cite{Excitonic_Ca2RuO4,PhysRevLett.116.017203,arxiv_2205_12123}.

 
\subsection{RIXS on \Na{}}

\begin{figure}[!t]
  \begin{center}
    \includegraphics{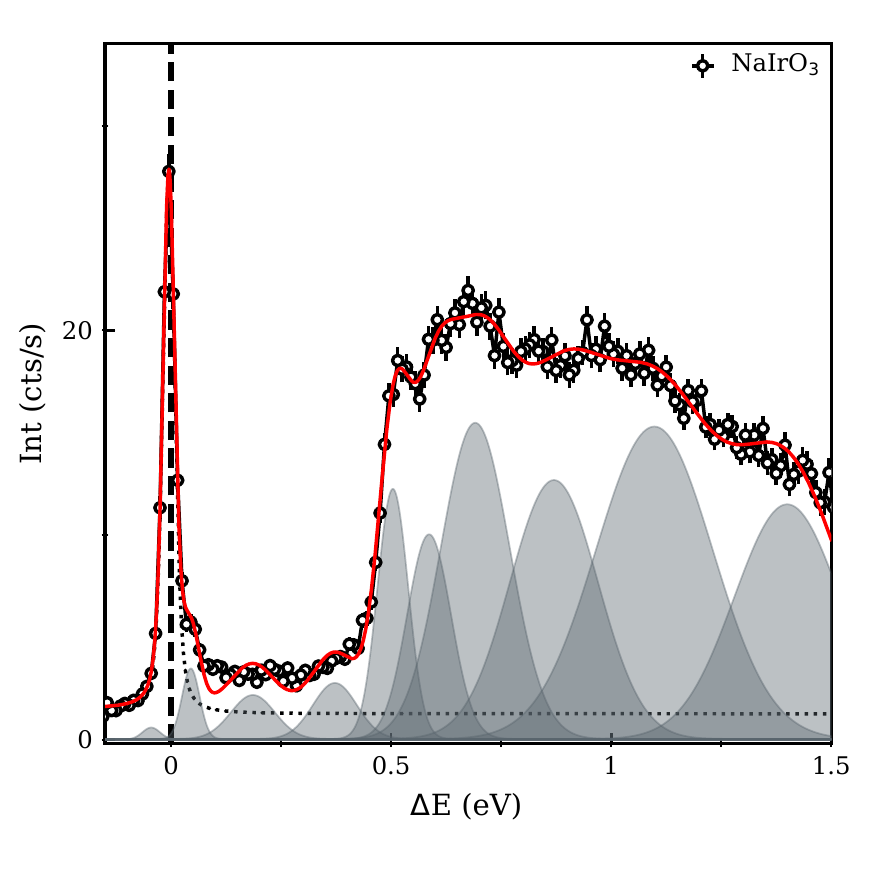}
  \caption{Ir L$_3$ RIXS intensity, $E_i \! = \!  11.217$ keV , in \Na{} at $T \!= \!20$~K. Vertical dotted line indicates \eloss{} $\!=\!  0$. Red line is a fit to a Voigt elastic line (dotted line) and a set of Gaussian peaks (grey shaded).}
  \label{fig:Na}
  \end{center}
\end{figure}

\begin{figure*}[!t]
  \begin{center}
    \includegraphics{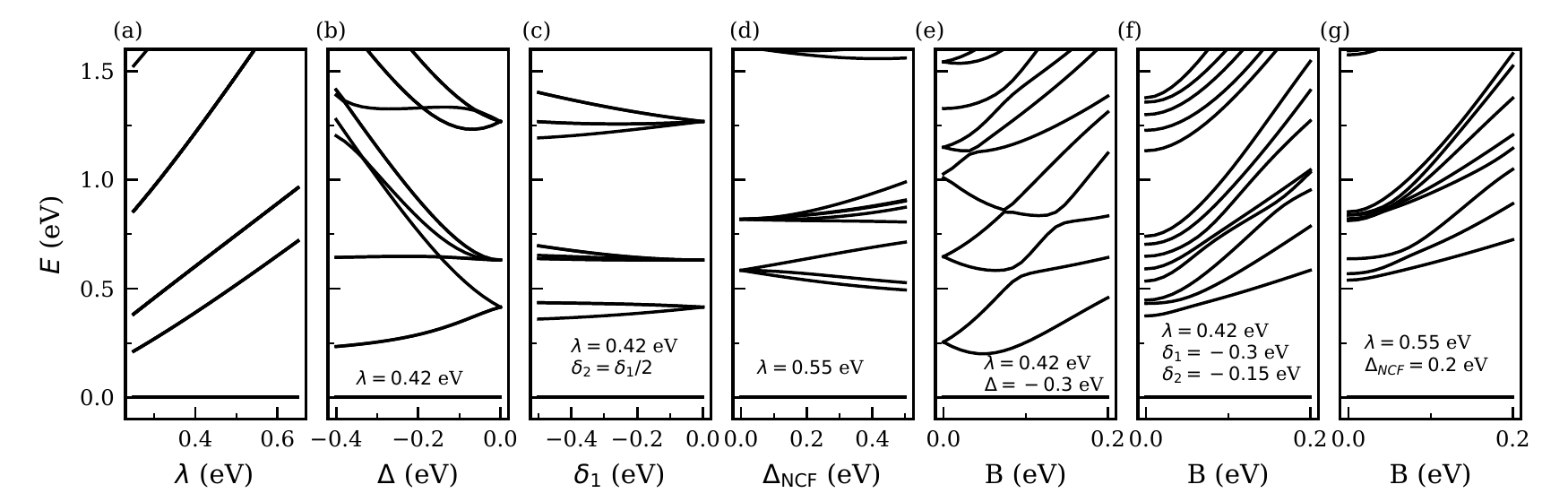}
 \caption{Excitation spectrum for a single Ir$^{5+}$ ions with $J_H \! =\!  0.285$~eV as a function of (a) atomic spin orbit coupling $\lambda$, (b) trigonal field, $\Delta$, (c), tetragonal field ,($\delta_1$, $\delta_2$), and (d) non-cubic fields ($\Delta_{NCF}$). We also show the excitation spectrum as a function of exchange field for fixed values of (e) trigonal, (f) tetragonal, and (g) non-cubic crystal fields.}
  \label{fig:TS_diagram}
  \end{center}
\end{figure*}

After having stablished the Ir$^{5+}$ oxidation state, we now bring our attention to the RIXS spectrum of \Na{}, shown in Fig. \ref{fig:Na}. There is a stark qualitative difference between the observed RIXS spectra for \Na{} and that of \SrCa{}. In contrast to other typical Ir$^{5+}$ iridates that have well-defined $\Delta E \approx 250$~meV RIXS features from inter-$t_{2g}$ split by 3/2 $\lambda$ excitations \cite{Bhowal_theory_double_perovskite_Ca_MMg,Breakdown_J0_band_structure,Bhowal_Hamiltonian_d4,Bhowal_2021,PhysRevMaterials.5.054604}, we observe a weak and broad inelastic signal that spans from  $\Delta E$ of 0.5~eV up to 1.5~eV and can be fit to a minimum of six Gaussian peaks. Below 0.5~eV there is a nearly constant small intensity above background, but no discernible peak. In the following, we discuss possible explanations for the observed RIXS spectrum.

\begin{figure}[!t]
  \begin{center}
    \includegraphics{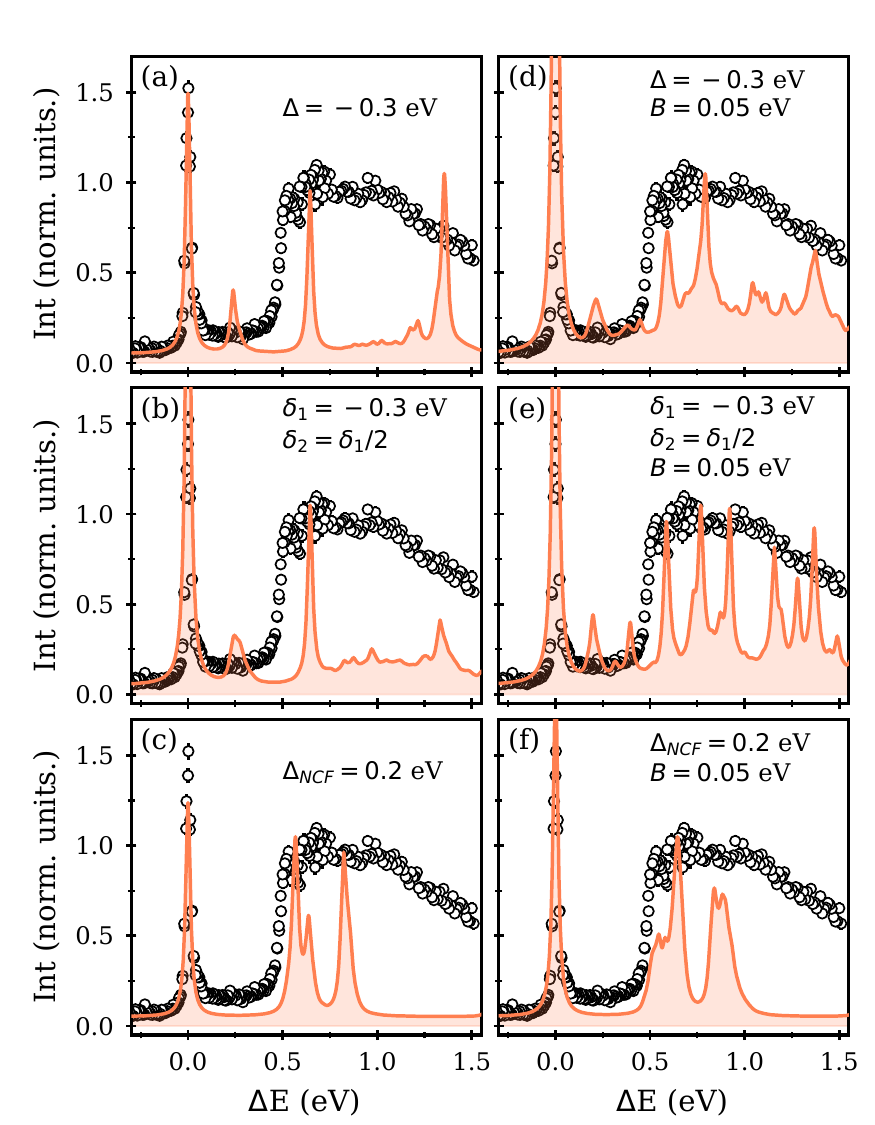}
  \caption{Simulated single ion RIXS spectrum for different crystal field environments, (a)- (c), and exchange splitting fields (d)-(e).}
  \label{fig:NaSingle}
  \end{center}
\end{figure}

We start by discussing the excitation spectrum of Ir$^{5+}$ within a single ion picture corresponding to the reported limit of a Mott insulating state in \Na{}. As shown in Fig. \ref{fig:TS_diagram} (a), in an undistorted octahedral environment with $J_H \! = \! 0.285$~eV two excited states are expected for $E < 0.5$ eV, an energy scale set by the spin-orbit coupling constant, $3\lambda/2 \! = \! 0.3- 0 .5$~eV \cite{Rau_AnnRevCondMatt}. The value of $J_H$, consistent with reported values for Ir \cite{PhysRevB.104.L100416,PhysRevB.95.235114}, is chosen to provide the best agreement between the data and the RIXS intensity calculations below. Trigonal ($\Delta$), tetragonal ($\delta_1,\delta_2$) and other non-cubic distortions ($\Delta_{NCF}$), act to further split the $t_{2g}$ states increasing the number of available excited states, as shown in Fig. \ref{fig:TS_diagram} (b), (d) and (f) for $\lambda \! = \! 0.42$~eV. We note that $\Delta_{NCF}$ parametrizes possible large non-cubic crystal fields resulting from low symmetry distortions of the IrO$_6$ octahedra as occurs in the post-perovskite NaIrO$_3$ ($\mu \!= \! 0.28 \mu_B/\rm{Ir}$, O-Ir-O $\approx \! 73.2 \deg$ and average $d_{\rm{Ir-Ir}} \! = \! 3.03 {\rm \AA}$). In this case, the $t_{2g}$ crystal field Hamiltonian includes the non-zero diagonal terms ($\Delta_{NCF}/2$,$\Delta_{NCF}$,$0$) \cite{Bhowal_2021}. In Fig. \ref{fig:NaSingle} (a)-(c) we show the calculated powder averaged RIXS  spectrum including a normal distribution of crystal field values to account for local ligand disorder with $\mu$ equal to $\Delta \! =\!  -0.3$~eV, $\delta_1 \! = \! -0.2$~eV with $\delta_2 \! = \! \delta_1/2.5$, and $\Delta_{NCF} \! = \! 0.2$~eV respectively, and $\sigma \! = \! 0.1$~eV which provide the best description of the RIXS spectrum. We set $\lambda \! = \! 0.42$~eV for all the calculations except those including $\Delta_{NCF}$ for which $\lambda \! = \! 0.55$~eV. For a better qualitative agreement with the data a linear background is added to the calculation. We find that these calculations fail to accurately describe the experimental data for any physically reasonable value of crystal fields.  

For all three models discussed, the magnitude of the crystal field that is  necessary to observe suppressed inelastic intensity in the region \eloss{}$<0.5$~eV and obtain a broad set of excitations above \eloss{}$>0.5$~eV leads to a $J\neq 0$ ground state: $\langle J^2_{\Delta} \rangle = 1.647$, $\langle J^2_{\delta_1,\delta_2} \rangle = 0.096$ and $\langle J^2_{\Delta_{NCF}} \rangle = 0.085$. Given the finite  $\langle J^2 \rangle$ values we find, particularly for the trigonally distorted scenario, the small and weakly temperature dependent magnetic susceptibility measured in \Na{} would only be consistent with these scenarios if the non-zero magnetic moments interact with a large Ising anisotropy that acts to freeze out thermal magnetic fluctuations and onset of long range magnetic order for T$\! < \! 300$~K. Magnetization measurements have not been reported above 300~K for this compound, so such a scenario is within the realm of possibility. To account for a possible magnetic order in \Na{},  we include an effective in-plane exchange splitting field $B$ in our RIXS model \cite{EDRIXS} as a way to simulate any magnetism-induced $t_{2g}$ splitting in \Na{} \cite{fujiwara_detecting_2021}, as shown in Fig.~\ref{fig:TS_diagram} (c),(c) and (g). We sample an uniform distribution with central value for $B\! =\! 0.05$ ~eV and $\sigma = 0.02$~eV to model local-disorder-induced modifications of magnetic interactions. The resulting spectrum shows decreased inelastic spectral weight for $\Delta E \! < \! 500$~meV and a broadening of the excitation continuum above $ \Delta E\! > \! 500$~meV, as shown in Fig. \ref{fig:NaSingle} (b) and (d). While these calculations produce a qualitative agreement with the data, a direct confirmation of T$ < 300$~K magnetic ordering in \Na{} is necessary to justify these model. We note that the effect of the exchange splitting in the case of non-cubic crystal field,  Fig. \ref{fig:NaSingle} (f), is negligible as the degeneracy of the $t_{2g}$ orbitals is already broken by the crystal field. Given the inadequacy of this model to describe the RIXS spectra above 1~eV, we believe it is not a likely scenario. 

Thermodynamic and transport measurements for \Na{} have so far been interpreted in terms of a $J\!=\!0$ Mott insulator, but the data are also generally consistent with expectations for a narrow gap semiconductor. Indeed, the non-magnetic insulating post-perovskite version of \Na{} ($\theta_W \! = \! - 2.2$~K, $\mu_{eff}\!=\!0.28$~$\mu_B$/Ir) \cite{BREMHOLM2011601} has been associated to a $S = 0$ band insulating state [Fig.~\ref{fig:Model} (a)] \cite{Breakdown_J0_band_structure,Nature_of_insulating_NaIrO3}. To test this possibility, we have carried density functional theory calculations including spin-orbit coupling (DFT+SOI) of the band structure of \Na{} as shown in Fig~\ref{fig:bands} (See Appendix \hyperref[sec:newsec]{A} for a non-SOI band structure calculation). These calculations were carried out using both the experimentally determined crystal structure as well as for a relaxed unit cell assuming perfect Na stacking along the c-axis. The starting and relaxed lattice parameters and atomic positions are shown in Appendix \hyperref[sec:newsec]{A}. We observe a large relaxation of the lattice and atomic positions, consistent with other reports for d$^4$ iridates \cite{PhysRevB.105.115123} and the post-perovskite version of \Na{} \cite{Breakdown_J0_band_structure}.

For both structures, our band structure calculations reveal nearly flat conduction and valence bands of mostly Ir $5d$ character [Fig.~\ref{fig:bands} (a)-(c)], as shown by our density of states calculations (DOS), shown in Fig.\ref{fig:bands} (d)-(f). The experimental structure yields a metallic state [Fig.~\ref{fig:bands} (a)], which is rendered insulating by the introduction of $U = 4$~eV, as shown in Fig.~\ref{fig:bands} (b). However, the band structure of \Na{} is sensitive to the lattice parameter and atomic positions, with the fully relaxed structure yielding a band insulator with $E_{gap}\! =\! 0.5$~eV across the Brillouin Zone (BZ) [Fig.~\ref{fig:bands} (c)]. This phenomenology resembles that of post-perovskite \Na{} in which DFT calculations for the relaxed structure find that the largely distorted oxygen octahedra lead to a 600~meV separation of the $t_{2g}$ states, comparable to Hunds coupling in iridates, thus stabilizing a non-magnetic $S \! =\! 0$ state. However, for the experimental structure, the electronic structure is found to be metallic with an indirect gap and electron-electron correlations, $U$ are necessary to obtain an insulating state \cite{Breakdown_J0_band_structure}.Additionally, LDA+Gutzwiller calculations find that the SOI is the key player in explaining the electronic groundstate of the post-perovskite \Na{}, such it has been argued to be a band insulator, rather than a correlated Mott insulator \cite{Nature_of_insulating_NaIrO3}. We therefor consider the possibility that honeycomb \Na{} is also a flat band insulator to provide an alternative explanation for the RIXS spectra. 

\begin{figure}[!t]
  \begin{center}
    \includegraphics{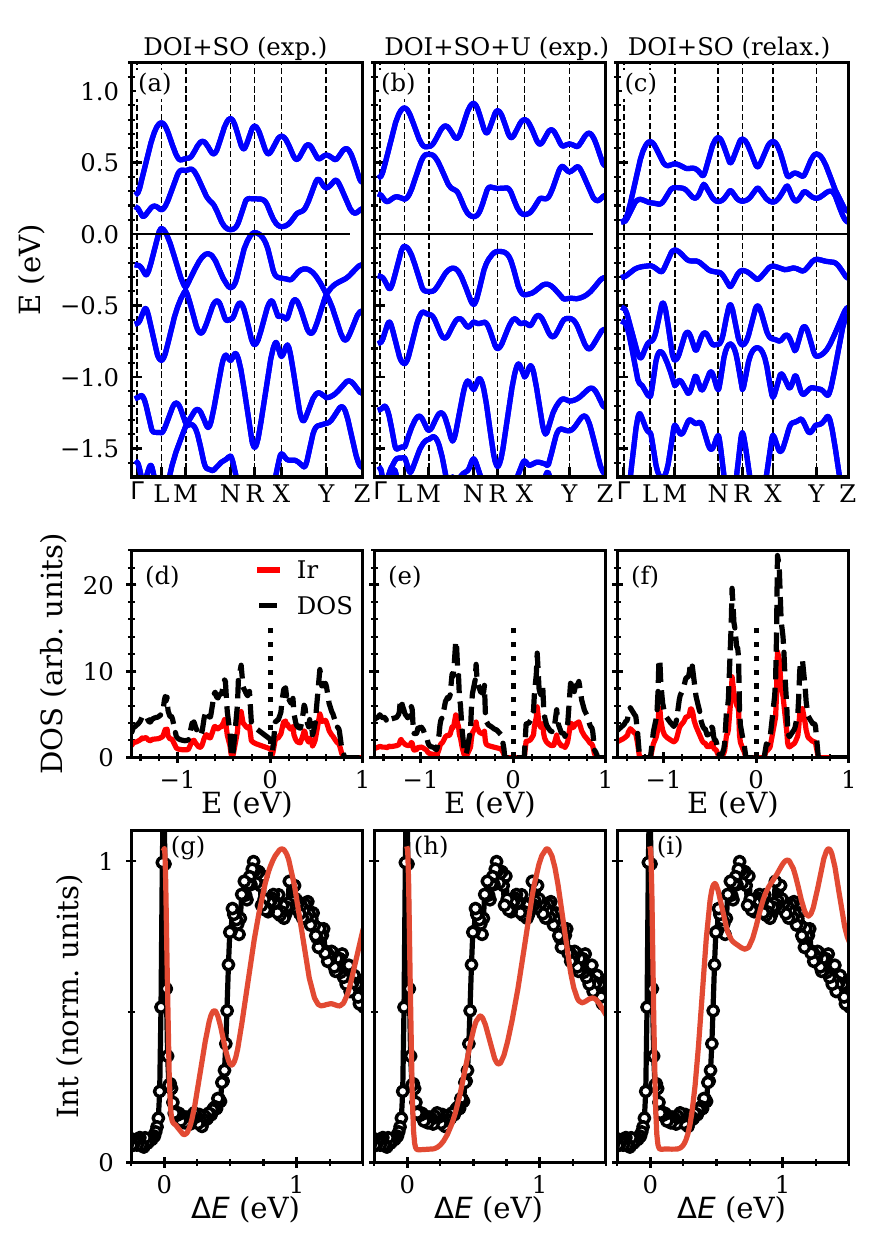}
  \caption{(a) DFT +SO band structure and (b) DFT+SO+U band structure, with $U \! = \! 4$~eV, of \Na{} for the reported strucutre. (c) DFT+SO band structure of \Na{} for the fully relaxed structure. The associated calculated total and Ir $5d$ density of states (DOS) are shown in (d)-(f). The resulting calculated RIXS spectrum compared to the measured RIXS spectrum at resonance for the reported structure is shown in panels (g)-(h)}
  \label{fig:bands}
  \end{center}
\end{figure}
In small gap insulating materials, RIXS can be regarded as an alternative technique to electron energy loss spectroscopy to resolve the unoccupied electronic structure. Due to the localized nature of the core hole, the energy and momentum dependence of the inelastic features can be related to joint density of states of the photoexcited electron and valence hole excitations across the gap, JDOS($Q,\Delta E$) \cite{Old_RIXS_eh_excitations_I,Old_RIXS_eh_excitations_II,PhysRevB.72.085221,1T-TiSe2_RIXS_eh_excitations}. We calculate the RIXS intensity as the JDOS between an occupied and unoccupied Ir states as a function of \eloss{} \cite{1T-TiSe2_RIXS_eh_excitations}. Given the powder nature of our samples, the low-symmetry unit cell of \Na{} and the three dimensional nature of the band structure of \Na{} the calculation of a complete momentum integrated JDOS(($Q,\Delta E$) over all the possible values of $Q$ is complex and computationally expensive. We perform an approximate calculation to the JDOS involved in the RIXS process, by fixing $Q = 0$ and integrating over interband transitions along the high-symmetry directions in the Brillouin zone ($\Gamma\!-\!L\!-\!M\!-\!N\!-\!R\!-\!X\!-\!Y\!-\!Z$). The resulting RIXS intensity, convolved with the experimental resolution, is shown in Fig.~\ref{fig:bands} (g)-(i) for both  $U \! =\! 0$, $U\! =\! 4$~eV, and relaxed calculations.  For the experimental structure, the calculated RIXS intensity is characterized by two broad peaks with central energy depending  on $U$. Neither calculation for $U \! =\! 0$, or $U \! =\! 4$~ eV provides an agreement with the observed RIXS dispersion. On the other hand, the calculated RIXS intensity for the relaxed structure displays a continuum of excitations that sharply onsets near \eloss{}$\!=\!500$~meV, in qualitative agreement with the \Na{} RIXS spectrum. While DFT calculations tend to underestimate the gap size and a more refined calculation of the electronic structure should appropriately treat $U$, this analysis is strongly suggestive of \Na{} being in a non-magnetic $S\! =\! 0$ state. In this case, spin-orbit coupling  generates a flat band insulating state and \Na{} may be a potentially interesting platform to study topological charge excitations if a suitable means to tune the chemical potential can be found \cite{PhysRevB.104.235202}.

\begin{figure}[!t]
  \begin{center}
    \includegraphics{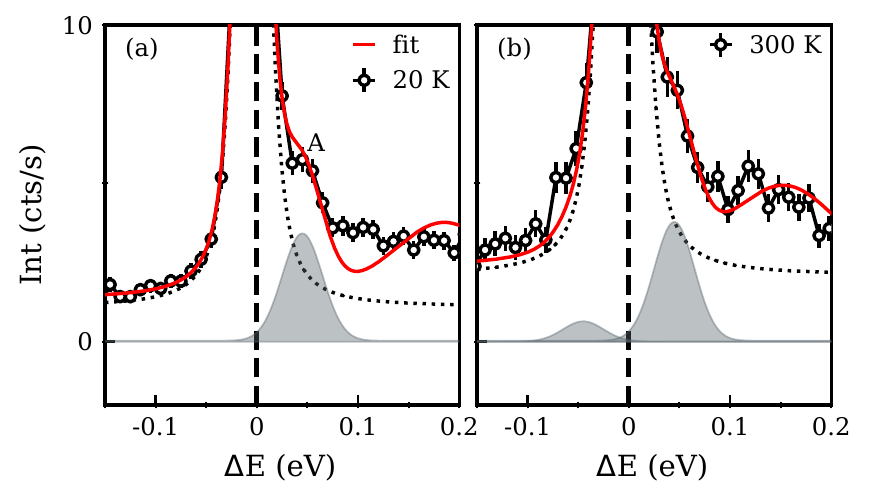}
\caption{ Ir L$_3$ RIXS intensity, $E_i = 11.217$ keV , in \Na{} at $T \!= \!20$~K (a) and $T \!= \!300$~K (b) in the range $E_\mathrm{loss} \!= \![ -0.15, 0.20]$~eV. Red line is a fit to the spectrum up to 1.5~eV as described in the main text. We highlight feature A, fitted to a resolution limited Gaussian corresponding to a low energy excitation. }
  \label{fig:Na_T}
  \end{center}
\end{figure}

In addition to the excitations above 500~meV, we find another low energy 45~meV excitation in the RIXS spectra of \Na{} that is not captured by either model. Figure~\ref{fig:Na_T} highlights the low-energy RIXS spectra around the elastic line. At $T\!=\!20$~K, we observe a clear shoulder on the elastic line that can be fit with a single Gaussian centered at an energy of 45~meV [Fig.~\ref{fig:Na_T} (a)]. This feature is also apparent in the thermally broadened elastic line at 300~K and its intrinsic origin is confirmed by the appearance of the corresponding excitation on the energy gain side, with an intensity following detailed balance [Fig.~\ref{fig:Na_T} (b)]. No charge excitations at 45~meV are predicted in either the large spin-orbit coupling  $J\!=\!0$ model, or the band insulating $S\!=\!0$ model. We associate this feature to scattering from phonons which, although expected to be minimal in our 90 degree scattering angle configuration, cannot be ruled out \cite{PhysRevLett.115.096404,PhysRevB.94.195131}. It is worth noticing that in the less likely scenario of a $J\!=\!0$ magnetic ground state in \Na{}, the Ir moments must then be magnetically ordered and interact through a strong Ising anisotropy to be consistent with the observed magnetic susceptibility. A lower bound on the magnetic interaction strength may be estimated by setting the Curie Temperature T$_c$ to the upper temperature limit of susceptibility measurements giving $J\! > \! 2/3 {\rm k}_B \times 300 \approx 16$~meV for a mean field Ising magnetic on the honeycomb lattice with weakly dispersing and gaped spin-waves appearing on an energy scale of at least $3J\approx 48$~meV, for $S=1$. 

\section{Conclusions}
In summary, we have harnessed the unique capability of RIXS to probe the local electronic structure of insulating compounds to identified two different electronic ground states in two newly synthesized powder Ir$^{5+}$ honeycomb iridates. The ground state of \SrCa{} is consistent with the proposed $J\! =\! 0$ singlet and constitutes the first realization of a non-magnetic honeycomb iridate. On the other hand, the RIXS spectra of \Na{} departs from that of other pentavalent iridates and cannot be explained within the single ion excitation spectrum. Based on DFT+SOI calculations we propose that the largely distorted local Ir environment leads to a band insulator $S \!=\!0$ ground state. We hope our results lead to future higher temperature magnetization required to confirm Ising magnetism or to further sample growth efforts to obtain single crystals in which to map the momentum dependence of the electron-hole excitations across the gap and rule out a possible Ising-like magnetic transition in \Na{}. Our results highlight the potential of RIXS to distinguish a Mott insulator from a band insulator in non-magnetic insulating materials in which thermodynamic measurements cannot provide such a distinction.

We acknowledge helpful conversations with Michael DiScala, Jung Ho Kim and Mark Dean. Work at Brown University was supported by the U.S. Department of Energy, Office of Science, Office of Basic Energy Sciences, under Award Number DE-SC002165. TMM and JC acknowledge support from the Institute for Quantum Matter, an Energy Frontier Research Center funded by the U.S. Department of Energy, Office of Science, Office of Basic Energy Sciences, under Award DE-SC0019331. Use of the Advanced Photon Source at Argonne National Laboratory was supported by the U. S. Department of Energy, Office of Science, Office of Basic Energy Sciences, under Contract No. DE-AC02-06CH11357.

\section{Appendix A}
\label{sec:newsec}

\begin{figure}[!]
  \begin{center}
    \includegraphics{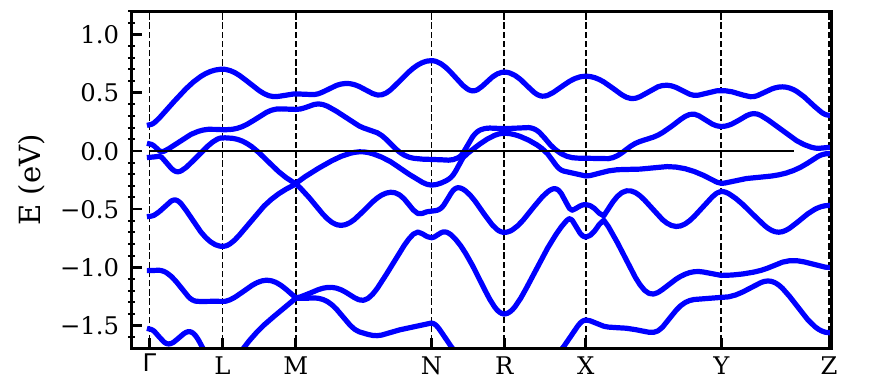}
\caption{Calculated band structure of \Na{} without SOI.}
  \label{fig:NoSOI}
  \end{center}
\end{figure}

In Fig.~\ref{fig:NoSOI} we show the calculated band structure of \Na{} without SOI. The resulting dispersion is metallic, highlighting the role of spin-orbit coupling in driving the band insulating state in \Na{}.

In Table~\ref{tab:atomp} we show the atomic positions for the experimental (($a$ = 5.281 $({\rm \AA})$, $b$ = 5.287 $({\rm \AA})$, $c$ = 6.00 $({\rm \AA})$, $\alpha$= 90.002$^{\circ}$, $\beta$= 115.7$^{\circ}$, $\gamma$= 60.11$^{\circ}$) and relaxed unit cell ($a$ = 5.509 $({\rm \AA})$, $b$ = 5.399 $({\rm \AA})$, $c$ = 6.708 $({\rm \AA})$, $\alpha$= 104.1$^{\circ}$, $\beta$= 121.27$^{\circ}$, $\gamma$= 59.39$^{\circ}$).

\begin{table}[!t]
    \centering
    \begin{tabular}{c|ccc|ccc}
      \hline \hline
      & \multicolumn{3}{c|}{Exp. Cell} & \multicolumn{3}{c}{Reduced Cell} \\
      \hline
      Atom & x & y & z & x & y & z\\
      \hline
      Ir & 0.164 & 0.156 & 0.010 & 0.170 & 0.162 &  0.001\\
      Na & 0.500 & 0.500 & 0.000 & 0.500 & 0.500 & 0.000\\
      Na & 0.500 & 0.500 & 0.500 & 0.500 & 0.500 & 0.000\\
      O & 0.948 & 0.468 & 0.153 & 0.987 & 0.493 & 0.189\\ 
      O & 0.239 & 0.844 & 0.206 & 0.236 & 0.870 & 0.182\\
      O & 0.639 & 0.103 & 0.190 & 0.613 & 0.117 & 0.182\\
      \hline \hline
    \end{tabular}
    \caption{Summary of the experimental and reduced atomic positions for the experimentally reported structure  and the fully relaxed cell.}
    \label{tab:atomp}
\end{table}

\vspace{100pt}
%

\end{document}